\documentclass[12pt,a4paper]{article}
\usepackage{amsmath}
\usepackage{graphicx}
\usepackage{times}
\begin{document}
\begin{titlepage}
\title{ The new scattering mode emerging\\ at the LHC?}
\author{ S.M. Troshin, N.E. Tyurin\\[1ex]
\small  \it SRC IHEP of NRC ``Kurchatov Institute''\\
\small  \it Protvino, 142281, Russian Federation}
\normalsize
\date{}
\maketitle
\begin{center}
\end{center}
\begin{abstract}
We point out  the several experimental manifestations of the new reflective scattering mode in elastic scattering and multiparticle production at the LHC energy range.
\end{abstract}

\end{titlepage}
\setcounter{page}{2}
\section*{Introduction}
Discussion of the geometry of the soft hadron processes is often focused on the impact--parameter dependence of the inelastic interactions since those are considered to be a driving force for the elastic scattering (shadowing). This geometry is determined by the nonperturbative QCD dynamics. 
It should be noted that  during  a long  time the elastic scattering  was considered to be consistent with the BEL picture, i.e. the picture where the interaction region becomes blacker,  edgier and larger \cite{valin1}. Such behavior corresponds  to the black--disk limit saturation due to the incoherent parton interactions  \cite{anis1}.  The model-based conclusions on the black-disk limit do not  exclude  an existence of another option for the hadron asymptotical dependence. 

 As it was noted on the model ground in \cite{edn}, the inelastic overlap function at the asymptotical energies can acquire a peripheral form in  the impact parameter representation.  The 
above-mentioned peripherality of the inelastic overlap function was treated as a manifestation of an emerging transparency in the central hadron collisions (or in vicinity of  impact parameter of the colliding particles $b=0$). Later on, the interpretation based on the peripheral form of the inelastic overlap function has been generalized and specified in papers \cite{bd1,bd2,bd3,degr} where such phenomenon  was related to antishadowing or reflection in the hadron interactions.   The  concept of the on-shell optical potential also leads to conclusion on the central grayness in the inelastic overlap function  \cite{arriola}.

A recent analysis \cite{alkin} of the elastic scattering data obtained by the TOTEM Collaboration at  
$\sqrt{s}=7$ TeV \cite{totem} has revealed an existence  of this novel feature in strong interaction dynamics due  to transition   to this scattering mode   referred nowadays also as resonant scattering \cite{anis}. Contrary to the black disk limit asymptotics, this new mode corresponds to the coherent parton interactions relevant for the confinement dynamics and is in agreement with the Chew and Frautschi conjecture  that the  strong interactions are to be ``as strong as possible'' \cite{chew,chew1}.

A gradual transition to the REL picture, i.e. picture when the interaction region becomes reflective (the term reflective means that the elastic scattering matrix element acquires negative values) at the center ($b=0$) and simultaneously being  edgier, larger and completely black in the ring at periphery, seems to be observed by the TOTEM  under the measurements of the $d\sigma/dt$ in elastic $pp$--scattering \cite{totem}. Several phenomenological models are able to reproduce such transition and among them the one  based on the rational unitarization of the leading vacuum Regge--pole contribution with intercept greater than unity 
\cite{edn}, and similar models   known under the generic name of the unitarized supercritical Pomeron (cf. \cite{anis} for a recent  discussion and the references).

There are two aspects of the asymptopia that can be emphasized  in connection with what has been said above. What is the limit for the partial or impact parameter amplitude reached at $s\to \infty$: the black disk limit of $1/2$ or the limit imposed by unitarity,  $1$. The second related issue  in case if the unitarity  saturation occurs: at what  energy value  the black limit $1/2$ is  crossed. The analysis performed in \cite{alkin} indicates that the black-disk limit has already been crossed at 
$\sqrt{s}=7$ TeV, while a simple extrapolation based on the gaussian impact parameter dependence of the elastic scattering amplitude  assumes the crossing happening at higher energies \cite{drem} but such  a dependence is at variance with large $-t$ data measured at $\sqrt{s}=7$ TeV. 

In this note we  address the above matters in the elastic scattering and multiparticle production processes at the LHC energy range. The aim is to indicate the experimental observables whose measurements could provide an information on the asymptotic mode.

\section{The experimental evidence for the reflective mode in elastic scattering} 
We are  using  here a common conjecture on the pure imaginary scattering amplitude\footnote{It should be noted that saturation of the black--disk limit or the unitarity limit leads to a vanishing real part of the scattering amplitude, $\mbox{Re} f\to 0$ in the region where the both limits are saturated \cite{tr}. The recent data \cite{tote} on the precise measurements of the ratio of the real to the imaginary part of the forward  amplitude are consistent with decreasing energy dependence of this ratio.} and perform the replacement $f\to if$ at high energies.
On the base of the  existing experimental trends, it seems natural to suppose a monotonic increase of the elastic scattering amplitude $f(s,b)$ with the energy without oscillations over $s$. 

  The analysis \cite{alkin}
 provides a clue that the unitarity limit and not a black disk one is saturated. This analysis deals with reconstruction of the elastic scattering amplitude in the impact parameter representation.  It should be noted that we are discussing asymptotic behavior of the impact-parameter dependent amplitude and not particular models for it.  
 It has been shown that the amplitude
$f(s,b)$ is greater than the black-disk limit of $1/2$ at $\sqrt{s}=7$ TeV at small impact parameters,  $f(s,b)=1/2[1+\alpha(s,b)]$, and the relative excess $\alpha$ being positive is still rather small at this energy in the region of the small $b$.  The value of $\alpha$ is about $0.08$  at $b=0$ \cite{alkin}. 
Therefore, the most relevant quantities to study crossing  the black-disk limit at the LHC energies are the elastic scattering amplitude $f(s,b)$ and the elastic overlap function $h_{el}(s,b)$, while the inelastic overlap function $h_{inel}(s,b)$) is less sensitive since its relative negative deviation at small values of $b$ is of the order of $\alpha^2$, i.e. $h_{inel}(s,b)=1/4[1-\alpha^2(s,b)]$, where $\alpha(s,b)$  does not vanish in the region $0\leq b<r(s)$ ($\alpha(s,b)=0$ at $b=r(s)$ and is negative at $b>r(s)$). The quantities $f$, $h_{el}$ and $h_{inel}$ enter the unitarity relation in the impact parameter representation:
\begin{equation}\label{un}
\mbox{Im}f(s,b)=h_{el}(s,b)+h_{inel}(s,b)
\end{equation}
and in the case under consideration it can be rewritten in the form:
\begin{equation}\label{un1}
h_{inel}(s,b)=f(s,b)[1-f(s,b)].
\end{equation} 

 To pinpoint the crossing of the black disk limit one should not be concentrated  on the integrated observables only since 
 those are much less informative and less sensitive than the differential ones. 
The analysis performed in \cite{alkin} is based on the consideration of the {\it differential} cross--sections since it is a most relevant method of the  study of the transverse plane interaction geometry.  Since the deviation $\alpha$ is small,  the effective reconfirmation of a conclusion on the black disk limit crossing requires  another  indications    in favor of  the black-disk limit overshooting (with the  subsequent unitarity limit saturation). 

Saturation of the unitarity limit corresponds to the limiting behavior $ S(s,b)\to -1$ at fixed $b$ and $s\to \infty$ and, by analogy with the reflection of light in optics,  can be referred as a pure reflective scattering  \cite{bd3}. 
The emergence of the reflective scattering is associated with increasing proton's density  with  the collisions' energy growth. It can be interpreted that far beyond some critical value of the density (corresponding to the black-disk limit) the colliding protons scatter  like the hard billiard balls. Such a behavior can be compared to  reflection of the incoming wave by a metal (changing phase of the incoming wave by $180^0$ due to presence of free electrons in a metal).  An increasing reflection ability emerging that way is associated with  a decreasing absorption according to the probability conservation expressed in the form of unitarity relation. 
The principal point of the reflective scattering  is  fulfillment of the inequalities  $1/2  < f(s,b) <1$ and $0  > S(s,b) > -1$ allowed by the unitarity relation \cite{bd1,bd2}. Our goal here is to discuss some of the experimental manifestations of this novel scattering regime. 
 
It was already noted that exceeding the  black-disk limit is a principal conclusion of the model--independent analysis of the impact--parameter dependencies of the amplitude performed in \cite{alkin}.  
The amplitude  at small $b$ values is most sensitive to the $t$--dependence of the scattering amplitude $F(s,t)$ in the region of large values of $-t$ (referred as deep--elastic scattering \cite{islam}).    In \cite{deepel} it was shown that the saturation of the unitarity limit corresponds to the relation
\begin{equation}\label{asmpt}
(d\sigma^{rfl}_{deepel}/dt)/(d\sigma^{abs}_{deepel}/dt)\simeq 4.
\end{equation}
Since at the LHC energy $\sqrt{s}=7$ TeV   the positive deviation from the black-disk limit at $b=0$ is small, the following
approximation is valid
\begin{equation}\label{asmpt1}
(d\sigma^{rfl}_{deepel}/dt)/(d\sigma^{abs}_{deepel}/dt)\sim 1+2\alpha(s,b=0).
\end{equation}

The absorptive models do not reproduce crossing of the black-disk limit  and provide a rather poor description of the LHC data in the deep--elastic scattering region (cf. e.g. \cite{b2,godiz,drufn}). 
Without fit to the data on the differential cross-section in this region of $-t$ the crossing of the black-disk limit can be easily missed.  This fact emphasizes again the  importance of consideration of the {\it unintegrated} quantities. In contrast to the absorptive models,
the Donnachie-Landshoff  model \cite{dln}, where the black disk limit is exceeded, is in a good agreement with the LHC experimental data on 
$d\sigma/dt$ at $\sqrt{s}=7$ TeV in the whole region of transferred momentum. 
This is also true for the models based on the rational or $U$--matrix form of unitarization. Fig. 1 
\begin{figure}[hbt]
\begin{center}
\hspace{0.5cm}
\resizebox{10cm}{!}{\includegraphics*{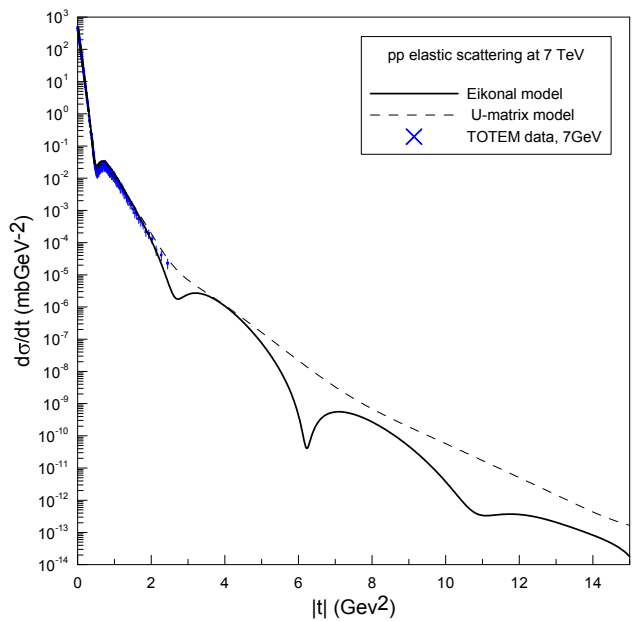}}
\end{center}
\vspace{-3.5cm}
\caption[ch1]{\small Description \cite{martyn} of differential cross--section of $pp$-scattering at $\sqrt{s}=7$ TeV with the use of absorptive (solid line) and reflective (dashed line) forms of unitarization.}
\end{figure}
 illustrates the above statements. Namely, the absorptive (eikonal)  models based on the exponential unitarization predict lower values for the differential cross-section in $pp$-scattering at $\sqrt{s}=7$ TeV and lead to appearance of the secondary bumps and dips  in the region of large  $-t$ values.

This important qualitative dissimilarity   of $d\sigma/dt$ in the deep-elastic scattering region originates  from the adopted forms of  the scattering amplitude in the impact parameter representation for the two unitarization schemes.
 \begin{figure}[hbt]
\begin{center}
\resizebox{12cm}{!}{\includegraphics*{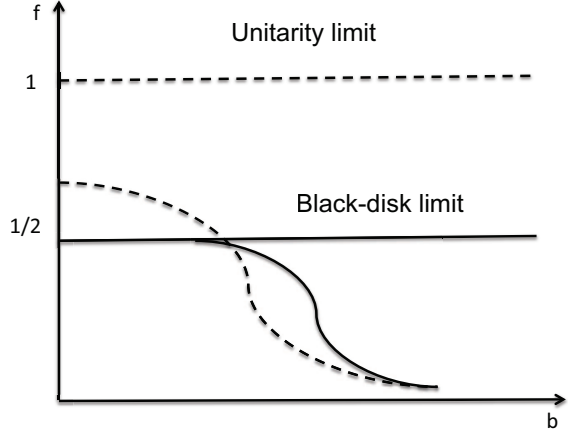}}
\end{center}
\vspace{-1cm}
\caption[ch1]{\small Schematic impact-parameter dependence of the amplitude $f(s,b)$ in the two cases of absorptive (solid line) and reflective (dashed line) forms of unitarization.}
\end{figure}
Concerning the amplitude 
\begin{equation}\label{stot}
f(s,b)=\frac{1}{4\pi}\frac{d\sigma_{tot}}{db^2}
\end{equation}
the following observations can be made:
\begin{itemize}
\item The models based on absorptive form of unitarization  have an almost flat $b$-dependence of $f(s,b)$ in the region of small and moderate impact parameter values since at the energy of $\sqrt{s}=7$ TeV the black-disk limit is believed as already reached at $b=0$  \cite{drufn} (cf. Fig.2). The result of such flatness is appearance (cf. \cite{halz}) of the secondary dips and bumps (solid line at Fig.~1). 

\item In an unitarized model (absorptive or reflective) the correct reproduction of  the experimentally observed total cross-section values requires the areas under the solid and dashed curves (cf. Fig. 2) to be equal. We consider that the dashed curve  corresponds to the experimental situation. Then, to reproduce   the observed total cross-section value, the solid curve (being almost flat at small $b$ since it cannot exceed $1/2$)  should be extended  to  larger values of the impact parameter, Fig. 2.  Since the slope parameter $B(s)$ \[B(s)\equiv \frac{d}{dt}\ln \frac{d\sigma}{dt}|_{-t=0}\]  is determined by an average value  $\langle b^2\rangle$, one can conclude that the observed value of $B(s)$ cannot not be well reproduced that way.  

The observed  excess above the black disk limit is not very significant at $\sqrt{s}=7$ TeV, but its increase is to be expected at higher energies. Thus the difficulties in the simultaneous description of the total cross-section and the slope parameter  in absorptive (eikonal) models would become more noticeable with the collision energy increase. 
\end{itemize}
\section{The clues for the reflective mode in multiparticle production} 
The most essential feature of the reflective scattering mode affecting observables in multiparticle production is the peripheral impact profile of the inelastic overlap function 
\begin{equation}\label{hin}
 h_{inel}(s,b)\equiv \frac{1}{4\pi}\frac{d\sigma_{inel}}{db^2} .
\end{equation}
However, this peripheral form will be reached at very high energies, and at the LHC energies the peripherality is not very distinctive. When the unitarity is saturated the transformation of the inelastic overlap function behavior is shown on Fig. 3.
\begin{figure}[h]
\begin{center}
\resizebox{8cm}{!}{\includegraphics*{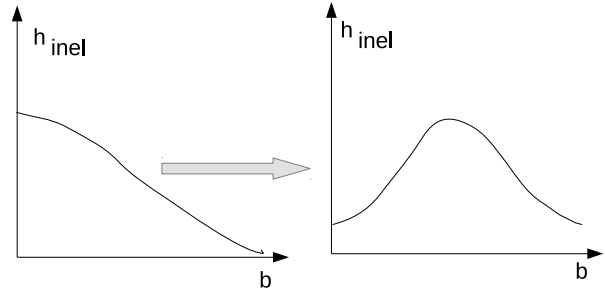}}
\end{center}
\caption[ch1]{\small Transformation of the inelastic overlap function with the energy increase from its initial central to peripheral form.}
\end{figure}
The total probability of an inelastic processes in the hadron collision at the energy $s$ and impact parameter $b$ is the following
\begin{equation}\label{pinel}
 P_{inel}(s,b)=4h_{inel}(s,b)= { \frac{d\sigma_{inel}}{2\pi bdb}},
\end{equation}
i.e.
\begin{equation}\label{inel1}
\sigma_{inel}(s)=2\pi\int_0^\infty  P_{inel}(s,b)bdb .
\end{equation}
Any observable, which desribes a multiparticle production process, $A(s,\xi)$ (where $\xi$ is a variable or a set of  variables), 
can be obtained from the corresponding impact-parameter dependent function $A(s,b,\xi)$
by integrating it with the weight function  $P_{inel}(s,b)$. 
Since this weight function has a prominent peripheral dependence  with a peak at $b=r(s)$,
at asymptotically high energies (cf. Fig. 3)
the following approximate relation is 
valid in the limit $s\to\infty$ \cite{ttint}:
\begin{equation}\label{mst}
A (s,\xi)\simeq A(s,b,\xi)|_{b=r(s)}.
 \end{equation}
The function $r(s)$ corresponds to the solution of the equation $ P_{inel}(s,b)=1$. Eq. (\ref{mst}) is applicable  at very high energies for such observables as mean multiplicity, average transverse momentum, anisotropic flows $v_n$ and multiplicity distribution $P_n(s)$. In general, this relation means that for the  production processes the relative range of  variations of the impact parameter 
 is decreasing with the energy and the most typical inelastic event at very high energy is the event with a non-zero value of the impact parameter in the region near $b=r(s)$ while the inelastic events at small and large impact parameter values should be suppressed. This suppression is strong at $s\to\infty$.
For the elastic scattering, the impact parameter profile is central  and expanding one with the energy.  
 
 We are going now to evaluate energy dependence of the average transverse momentum of produced particles at the asymptotic energies
 using Eq. (\ref{mst}).  
  The  geometrical picture of hadron collisions discussed earlier (cf. \cite{multrev}) supposes that at high energies and non-zero impact parameters
 the constituent quarks produced in the overlap region under the hadron collisions carry  
 large orbital angular momentum.
 It  has been estimated
as 
\begin{equation}\label{l}
 L(s,b) \propto  b \frac{\sqrt{s}}{2}D_C(b),
\end{equation}
where $D_C(b)\equiv D_C^1\otimes D_C^2$ is a convolution of the peripheral condensates' distributions in the two colliding
hadrons.
Due to preasumed  strong interaction
between the quarks this orbital angular momentum should lead to a coherent rotation
of the  overlap region as a whole.
This rotation is similar to rotation of a liquid which is considered to be a  quark-pion liquid \cite{multrev}.
This collective coherent rotation is considered to be an only source of the secondary particles' transverse momentum at
$s\to \infty$. 
Namely, it is  assumed that the rotation of transient matter is
affecting average transverse momentum of the secondary hadrons in proton-proton collisions.
 The following relation can be used 
\begin{equation} \label{ptl}
\langle p_T\rangle(s,b)=\kappa L(s,b),
\end{equation}
 where $L(s,b)$ is given by Eq. (\ref{l}) and $\kappa$ is a constant (it has  dimension of inverse length),
which can be related to the inverse hadron radius.
The average transverse momentum $\langle p_T(s)\rangle$ at $s\to \infty$ can be calculated then using Eq. (\ref{mst}) if one assumes 
that the condensate distributions in the hadrons are controlled by the pion mass, i.e.
\[
D_C^1\otimes D_C^2\sim \exp (-m_\pi b).
\]

With the model described in \cite{multrev} for the average transverse momentum $\langle p_T \rangle (s)$ at asymptotically high energies
we will have
\begin{equation}\label{apt}
\langle p_T \rangle (s) = cs^{\delta_C}\ln s ,
\end{equation}
where $\delta_C=1/2-{m_\pi}/{m_Q}$ and $m_Q$ is the mass of the constituent quark $Q$. Using  the value $m_Q\simeq 0.35$ GeV
one will have  for the exponent $\delta_C\simeq 0.1$. Of course, Eq. (\ref{apt}) is valid at $s\to \infty$.
The experimental data up to the energy $\sqrt{s}=2.36$ TeV can be well described  using the fit
\begin{equation}\label{apte}
\langle p_T \rangle (s) = a+cs^{\tilde\delta_C}
\end{equation}
 with the parameters
$a=0.337$ GeV/c, $c=6.52\cdot 10^{-3}$ GeV/c and $\tilde\delta_C=0.207$.
Eq. (\ref{apte}) is in a good agreement with the experimental data at $\sqrt{s}\leq 2.36$ TeV (cf. Fig. 4).
\begin{figure}[h]
\begin{center}
\resizebox{8cm}{!}{\includegraphics*{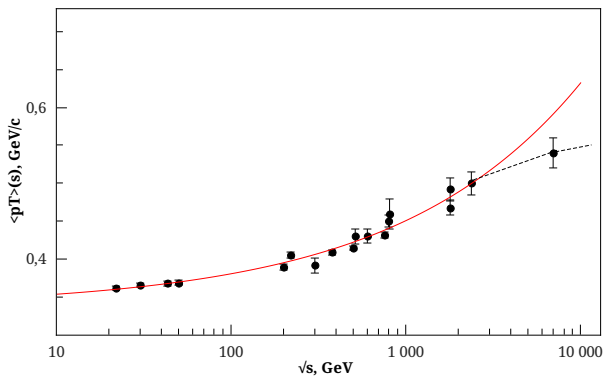}}
\end{center}
\caption[ch2]{\small Slow-down of an  energy dependence of the average transverse momentum.}
\end{figure}
As it is evident from Fig. 4, one might expect a slow-down of the  increase of the average transverse momentum, the point 
at $\sqrt{s}=7$ TeV lies below the respective extrapolation. Such slow-down can be interpreted as a beginning of transition to the asymptotic dependence Eq. (\ref{apt}) which has a lower value of the exponent. Thus, this slow-down then can be correlated with transition to the reflective scattering mode
and serve as its clue. Reflective scattering should also lead
to slow-down of the average multiplicity energy dependence \cite{ttphl}. The presence of the reflective scattering mode would lead to the simultaneous slow-down of the average transverse momentum and average multiplicity.
It should also be noted that multiplicity distribution would have a form of the product of the two stochastic distributions at the asymptotic energies if one adopt the Chou and Yang geometric picture of multiparticle production \cite{chyn}, i.e. the product of the two Poisson distribution one in $n_F$ and another one in $n_B$ where $n_F$ and $n_B$ are the multiplicities of the secondary particles in forward and backward hemispheres. The average multiplicity for both distributions would be 
\[
\langle n \rangle (s,b)|_{b=r(s)}/2.
\]
 In the model \cite{multrev}
it has power-like energy dependence at $s \to \infty$, i.e.
\begin{equation}\label{amul}
\langle n \rangle (s) = cs^\delta,
\end{equation}
where $\delta=\delta_C=1/2-{m_\pi}/{m_Q}\simeq 0.1$.
Thus, the transition with energy from the nonstochastic to  stochastic multiplicity distribution can be envisaged, in the case of the unitarity limit saturation since the product of two Poisson distributions gives another Poisson distribution multiplied by the binominal factor \cite{chyn}. But, the Chou-Yang approach would provide nonstochastic multiplicity distribution in case of the black limit saturation.

Another interpretation there was proposed in \cite{decoh}. 
Namely, the energy dependence of the average transverse momentum which has started to change in the
 region $\sqrt{s}>3$ TeV 
has been related to an appearance of the decoherence in the proton interactions which in its turn might results from the gradual 
phase transition of the strongly 
interacting transient state (quark-pion liquid) into the weakly interacting gas of quarks and gluons (quark-gluon plasma). 

Conclusion on the validity of the particular interpretation could be done on base of the average multiplicity measurements. If there is no simultaneous changes in the energy dependencies of $\langle p_T \rangle (s)$ and $\langle n \rangle (s)$  (when average transverse momentum changes its energy dependence but average multiplicity does not) one should choose the option based on the  emergence of the decoherence in the proton interactions.  Further experimental
studies at higher energies are necessary to make a choice between  the two above options.
\section*{Conclusion}
Thus, the two above discussed alternatives---saturation of the black disk limit or saturation of the unitarity limit---generate different mechanisms of the total cross--section growth at the energies $\sqrt{s}>7$ TeV. Namely, saturation of the black disk limit assumes the growth of the total cross--section due to increasing effective impact parameter values only, while the growth of the total cross-section can, in fact, be due to both factors --- increase of impact parameter values combined with an increase of the elastic scattering amplitude $f(s,b)$ with the energy at fixed $b$.   Therefore, the measurements of the elastic scattering at $\sqrt{s}=13$ TeV especially in the region of large $-t$  could be quite decisive 
regarding the  asymptotic picture and would help to determine which  scattering picture ---absorptive or reflective --- should be expected at the asymptotical energies. 

The saturation of the black disk limit and saturation of the unitarity limit lead to the qualitatively different multiplicity distributions at $s\to\infty$.
The measurements of the average transverse momentum, average multiplicity, multiplicity distribution, correlations of multiplicities in the forward and backward hemispheres as well as the other particle correlations related, in particular, the ridge and double--ridge effects  also could be useful in the searches for further possible experimental traces of the reflective scattering mode in hadron interactions. 
\section*{Acknowledgements}
We are grateful to E. Martynov for the granted permission to use the results  of ref. \cite{martyn} and  interesting discussions on the subject.

\end{document}